\documentclass[prb,showpacs,preprintnumbers,amsmath,amssymb,twocolumn,superscriptaddress]{revtex4-1}

\usepackage[dvipdfmx]{graphicx}
\usepackage{dcolumn}
\usepackage{bm}
\usepackage{color}

\usepackage{amsmath}	
\usepackage{amsfonts}
\usepackage{braket}

\begin{document}

\title{ 
Robust orbital diamagnetism of correlated Dirac fermions \\
in chiral Ising universality class
}

\author{Yasuhiro Tada}
\email[]{ytada@hiroshima-u.ac.jp}
\affiliation{
Quantum Matter Program, Graduate School of Advanced Science and Engineering, Hiroshima University,
Higashihiroshima, Hiroshima 739-8530, Japan}
\affiliation{Institute for Solid State Physics, University of Tokyo, Kashiwa 277-8581, Japan}

\begin{abstract}
We study orbital diamagnetism at zero temperature in $(2+1)$-dimensional Dirac fermions with a short-range interaction
which exhibits a quantum phase transition to a charge density wave (CDW) phase.
We introduce orbital magnetic fields into spinless Dirac fermions on the $\pi$-flux square lattice,
and analyze them by using infinite density matrix renormalization group.
It is found that the diamagnetism remains intact in the Dirac semimetal regime
as a result of a non-trivial competition between the enhanced Fermi velocity and magnetic-field-induced mass gap, 
while it is monotonically suppressed in the CDW regime.
Around the quantum critical point (QCP) of the CDW phase transition,
we find a scaling behavior of the diamagnetism characteristic of the chiral Ising universality class.
This defines a universal behavior of orbital diamagnetism in correlated Dirac fermions around a QCP,
and therefore the robust diamagnetism in the semimetal regime is a universal property of Dirac systems
whose criticality belongs to the chiral Ising universality class.
The scaling behavior may also be regarded as a quantum, magnetic analogue of the critical Casimir effect which has been 
widely studied for classical phase transitions.
\end{abstract}

\maketitle

\section{introduction}
Orbital diamagnetism of conduction electrons 
is a fundamemtal property of a material. 
Intuitively, it arises through the Lorentz force
acting on electrons' kinetic motions 
and therefore it is susceptible to the band structure of the system considered.
Especially in a semimetal with linear dispersions,
the Landau level structure is qualitatively different
from that in a conventional parabolic band system.
This leads to anomalous magnetic responses in Dirac semimetals, and their
orbital magnetic moment $M$ shows extremely strong diamagnetism with 
non-analytic dependence on the magnetic field at zero temperature, $M\sim -\sqrt{B}$ in two spatial dimensions.
This is much stronger than that in conventional metals, 
$M\sim -B$, for small magnetic fields.
Extensive theoretical studies have been done mainly for non-interacting Dirac systems
~\cite{McClure1956,Nersesyan1989,Ghosal2007,Koshino2011JP,Fukuyama2012,
Graphene_exp2015,Raoux2014,Gomez2011,Fukuyama2007,Koshino2010,Koshino2007,Koshino2011,
Sheehy2007,Principi2010,Yan2017} even in a mathematically rigorous manner~\cite{Savoie2012},
and various properties of diamagnetism 
have been theoretically discussed such as finite temperature effects~\cite{Graphene_exp2015}, 
roles of Berry phase~\cite{Raoux2014},
lattice effects,~\cite{Gomez2011},
effects of an elastic life time~\cite{Fukuyama2007},
effects of a non-zero gap~\cite{Koshino2010}, 
disorder effects~\cite{Koshino2007},
and weak interaction effects~\cite{Sheehy2007,Principi2010,Yan2017}.
In a realistic finite size sample with surfaces, an edge current will flow along the sample surface and
generate orbital diamagnetism, where net edge currents are generally robust to surface conditions
~\cite{Koshino2011,Kubo1964,Ohtaka1973,MMP1988,Tada2015}.
Experimentally, strong diamagnetism has indeed been observed in several systems such as graphene and bismuth,
and they are well understood based on free electron models as a direct consequence of the Dirac
band structures~\cite{Graphene_exp2015,Bi1934,Fukuyama1970,Fuseya2015}.
Furthermore,
the origin of diamagnetim has been identified as orbital contributions in 
Sr$_3$PbO~\cite{Suetsugu2021} and Bi$_{1-x}$Sb$_x$~\cite{Watanabe2021}.

Recently, there have emerged a variety of {\it strongly interacting} Dirac electron compounds
such as molecular crystal $\alpha$-(BEDT-TTF)$_2$I$_3$~\cite{Hirata2017}, 
magnetic layered system EuMnBi$_2$~\cite{Masuda2016},
perovskite oxides Ca(Sr)IrO$_3$~\cite{CaIrO2019}, and twisted bilayer graphene~\cite{TBG2018}.
Given these experimental developments,
it is natural to ask how the orbital diamagnetism behaves in a correlated Dirac system.
According to the previous theoretical study for graphene with the long-range Coulomb interaction~\cite{Sheehy2007}, 
the orbital diamagnetization $M$ is enhanced if one takes into account the Fermi velocity ($v_F$) renormalization
since $M$ is proportional to $v_F$ in the Dirac semimetal phase.
However, it is known that an external magnetic field induces a Dirac mass in presence of an electron interaction,
which is known as magnetic catalysis~\cite{Shovkovy2013book,Miransky2015review,Fukushima2019review,
Gusynin1994,Gusynin1996NPB}.
The Dirac mass generally suppresses diamagnetism and it competes with an enhancement of the Fermi velocity.
A perturbation study for graphene suggests that orbital magnetization is suppressed at zero temperature
as a result of the non-trivial competition of these two opposite effects~\cite{Yan2017}.

Suppression of diamagnetism may occur also in other related systems where 
effects of mass generations due to the magnetic catalysis are stronger than those of 
Fermi velocity renormalizations.
There are several kinds of magnetic catalysis corresponding to distinct types of field-induced orders,
such as antiferromagnetism, superconductivity, and charge density wave (CDW) order.
The critical behaviors around a quantum critical point (QCP) of the semimetal-insulator phase transition
have been well established mainly 
in absence of a magnetic field~\cite{QCP_Dirac,Sorella1992,Assaad2013,Otsuka2016,Wang2014,Li2015,
Hesselmann2016,Toldin2015,Corboz2018,Rosenstein1993,Wetterich2001,Herbut2006,Herbut2009,Ihrig2018}, 
and quantum criticality of magnetic catalysis can also be understood in a similar manner~\cite{Tada2020}.
The scaling analysis shows that
the Fermi velocity $v_F$ remains regular around a QCP~\cite{Igor2009vF}, but
numerical calculations demonstrate that 
$v_F$ decreases to some extent in interacting Dirac fermions which exhibit antiferromagnetic 
quantum phase transitions~\cite{Tang2018,Hesselmann_com2019,Tang_rep2019,Lang2019} and furthermore
the reduced $v_F$ was observed in the molecular compound $\alpha$-(BEDT-TTF)$_2$I$_3$~\cite{Unozawa2020}.
Therefore, it is natural to expect suppression of diamagnetism in these systems similarly to the 
long-range Coulomb interacting graphene~\cite{Yan2017}.
However, it is not clear whether or not diamagnetism generally gets suppresed also in other interacting Dirac systems.

In this work, we study orbital diamagnetism in a representative model of interacting Dirac fermions
exhibiting a CDW order
by unbiased numerical calculations with the infinite density matrix renormalization group (iDMRG)
~\cite{White1992,DMRG_review1,DMRG_review2,DMRG_review3,TenPy1,TenPy2}.
In this system, the Fermi velocity increases in presence of the interaction~\cite{Schuler2021},
and therefore it is a promising candidate system to realize robust diamagnetism.
Indeed,
we demonstrate that the orbital diamagnetization remains intact for weak interactions in the Dirac semimetal
regime, while it
monotonically decreases as the interaction strength is increased in the insulating regime. 
Furthermore, 
the orbital magnetization $M$ exhibits a universal scaling behavior
near the QCP, 
and the robust orbital magnetization is characterized as a universal property of Dirac systems 
whose criticality belongs to the
chiral Ising universality class.
Besides,
the scaling behavior of $M$ is analogous to a seemingly unrelated
phenomenon, the critical Casimir effect which has been extensively studied for classical phase transitions.
Our study would provide a fundamental understanding of the orbital diamagnetism in correlated 
Dirac fermions based on the quantum critical scaling.

\section{Model}
\begin{figure}[tbh]
\includegraphics[width=4.0cm]{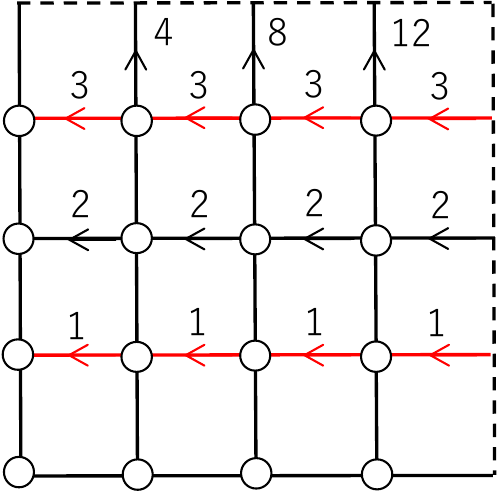}
\caption{The string gauge for a $L_x'=L_y=4$ system where white circles represent lattice sites and
the periodic boundary condition has been imposed.
The black (red) bond corresponds to the hopping $-t (+t)$.
Each number on the bonds corresponds to $A_{ij}$ in unit of $\delta B=2\pi n/L_x'L_y$.
}
\label{fig:string}
\end{figure}

We consider the $t$-$V$ model for
spinless fermions on a $\pi$-flux square lattice (also called staggered fermions) 
at half-filling under a uniform magnetic field,
which is one of the simplest realizations of interacting Dirac fermions similarly to the honeycomb lattice model
~\cite{QCP_Dirac,Sorella1992,Assaad2013,Otsuka2016,Wang2014,Li2015,
Hesselmann2016,Toldin2015,Corboz2018,Tada2019,Tada2020}.
The model has two Dirac cones in the Brillouin zone corresponding to
four component Dirac fermions in total.
The magnetization arises only from the electron orbital motion since there is no spin degrees of freedom,
which enables us to directly study the orbital magnetism.
The Hamiltonian is given by
\begin{align}
H=-\sum_{\langle i,j\rangle}t_{ij}c^{\dagger}_ic_j+V\sum_{\langle i,j\rangle}n_in_j,
\label{eq:H}
\end{align}
where $\langle i,j\rangle$ is a pair of the nearest neibghbor sites.
The hopping $t_{ij}=t_{ij}^{(0)}\exp (iA_{ij})$ contains  
the vector potential in the string gauge as shown in Fig.~\ref{fig:string}
corresponding to an applied uniform magnetic field~\cite{Hatsugai1999,Tada2020},
where $t_{j+\hat{x},j}^{(0)}=t\exp(i\pi y_j)$ and $t_{j+\hat{y},j}^{(0)}=t$ corresponding to the $\pi$-flux lattice. 
In the iDMRG calculation, the system is an infinite cylinder whose size is $L_x\times L_y=\infty\times L_y$
with the periodic boundary condition for the $y$-direction,
and we introduce superlattice unit cells with the size $L_x'\times L_y$.
The magnetic field is assumed to be spatially uniform and is an integer multiple 
of the unit value allowed by the superlattice size, 
$B=n\times \delta B \quad (n=0,1,2,\cdots, L_x'L_y)$ where 
$\delta B=2\pi/L_x'L_y$.
We consider two different system sizes $L_y=6,10$ to discuss finite size effects,
and typically use $L_x'=20$ for $L_y=6$ and $L_x'=10$ for $L_y=10$.
It turns out that the system can be regarded as a two dimensional system when the magnetic length
$l_B=1/\sqrt{B}$ is effectively shorter than the system size $L_y$~\cite{Tada2020},
which enables us to study (2+1)-dimensional physics by iDMRG.
We also simulate a system with $L_y=14$ only at zero magnetic field, which 
will be touched on at Sec.~\ref{sec:summary}.  
The system size in the present study is rather limited, but we will demonstrate that our results are 
consistent with those obtained in the previous studies for larger system sizes at $B=0$
~\cite{QCP_Dirac,Sorella1992,Assaad2013,Otsuka2016,Wang2014,Li2015,
Hesselmann2016,Toldin2015,Corboz2018}.
This consistency supports our discussions on the system in presence of the magnetic field which is less understood.
In this study, we use the open source code TeNPy~\cite{TenPy1,TenPy2}.

In absence of a magnetic field,
the Hamiltonian has sublattice ${\mathbb Z}_2$ symmetry which is related to the
chiral symmetry at low energy.
This symmetry is spontaneously broken for large interactions above the critical strength, $V>V_c=1.30t$,
and a charge density wave (CDW) state is realized where Dirac fermions acquire a dynamical mass
~\cite{Wang2014,Li2015,
Hesselmann2016,Tada2020}.
The CDW order parameter in the ground state has been discussed previously,
and it was shown that the CDW state is stabilized for any small $V>0$ in presence of the magnetic field $B\neq 0$
when the system size $L_y$ is large enough~\cite{Tada2020},
which is called the magnetic catalysis~\cite{Shovkovy2013book,Miransky2015review,Fukushima2019review,
Gusynin1994,Gusynin1996NPB}.
Especially near the QCP, $V\simeq V_c$, the CDW order parameter behaves as $M_{\rm CDW}\sim 
l_B^{-\beta/\nu}\sim B^{\beta/2\nu}$ with $\beta\simeq 0.54, \nu\simeq 0.80$
corresponding to the $N=4$ chiral Ising universality class.
Note that, if present,
the long-range part of the Coulomb interaction
would be less important around the QCP and would not affect the
criticality at zero magnetic field~\cite{Herbut2009,Toldin2015,Tang2018,Hesselmann_com2019,Tang_rep2019}.

Because the system is gapped at $B\neq0$ for any $V>0$ due to the magnetic catalysis,
the iDMRG numerical calculations with 
finite bond dimensions $\chi$ are stable and extrapolation $\chi\to\infty$ works well.
In the present study, we extrapolate the calculated ground state energy density at finite $\chi\leq 1600$
to $\chi\to\infty$
to obtain the true ground state energy density $\varepsilon$ for each set of $V,B$, and $L_y$.
Details of the extrapolation are discussed in Appendix ~\ref{app}.
All the numerical results in the following discussion are extrapolated ones.

\section{Numerical results}
Firstly, we breifly exlpain qualitative behaviors of the ground state energy density $\varepsilon$ 
in simple limiting cases before
discussing numerical results of the iDMRG calculations.
In the free Dirac fermions with a linear dispersion, single-particle energies are $\epsilon\sim l_B^{-1}$
with degeneracy $\sim l_B^{-2}$, which leads to the ground state energy 
$\varepsilon(B)-\varepsilon(0) \sim l_B^{-3}=B^{3/2}$.
The $B$-dependence becomes weaker in the deep CDW state with Dirac mass,
$\epsilon\sim ({\rm mass})+l_B^{-2}$ and hence $\varepsilon(B)-\varepsilon(0) \sim l_B^{-4}=B^2$.
These qualitative behaviors should hold not only deep inside each phase 
but also in general interaction strength in the phases.
Besides, the low energy Lorentz symmetry of the Dirac semimetal phase is kept up to $V=V_c$
~\cite{QCP_Dirac,Sorella1992,Assaad2013,Otsuka2016,Wang2014,Li2015,
Hesselmann2016,Toldin2015,Corboz2018}
and $\varepsilon(B)-\varepsilon(0) \sim l_B^{-3}$ holds also at the QCP~\cite{Tada2020}.
This scaling will be confimed later.

\begin{figure}[tbh]
\includegraphics[width=6.5cm]{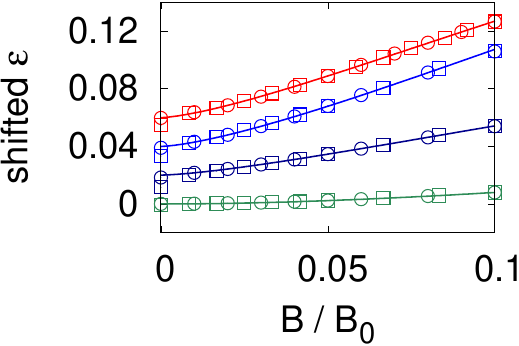}
\caption{The shifted ground state energy density $\varepsilon(B)$ for 
$L_y=6$ (squares) and $L_y=10$ (circles)
with the fitting curves $\varepsilon_{\rm fit}(B)$ (solid curves).
The interaction is $V/t=0, 0.5, 1.3, 2.0$ from the top to the bottom.
}
\label{fig:eB}
\end{figure}

Now we show the ground state energy density $\varepsilon(V,B)$ as a function of 
the magnetic field $B$
calculated by iDMRG with 
extrapolation $\chi\to\infty$ in Fig.~\ref{fig:eB}, where $\varepsilon(V,B=0)$ has been shifted for the eyes.
(The results $\varepsilon(V=0)$ have been simply obtained by direct diagonalization of
the non-interacting Hamiltonian for sufficiently long cylinder geometry.)
We see that the results for two different system sizes $L_y=6,10$ coincide
for relatively large magnetic fields $B\gtrsim 0.02B_0$ where
the magnetic length $l_B$ is effectively shorter than $L_y$,
although there are some deviations for small magnetic fields $B\lesssim 0.02B_0$ with longer $l_B$.
Therefore, finite system size effects are negligible as long as the magnetic length is effectively shorter
than the system size $L_y$ as previously mentioned. 
This means that our system is essentially two-dimensional with the size $\sim l_B\times l_B$.
In this scheme, we focus only on the magnetic length effectively shorter than $L_y=6$. 
It may seem
difficult to discuss (2+1)-dimensional physics because $l_B<L_y=6$ is too small, and in general, 
$L_y=6, 10$ would not be sufficient to discuss the true two dimensionality.
However, as was demonstrated in the previous study for the magnetic catalysis~\cite{Tada2020},
it is indeed possible to investigate (2+1)-dimensional physics with this range of magnetic
fields and the resulting physical quantities are consistent with those obtained by the previous studies
for larger system sizes at $B=0$~\cite{Shovkovy2013book,Miransky2015review,Fukushima2019review,
Gusynin1994,Gusynin1996NPB, QCP_Dirac,Sorella1992,Assaad2013,Otsuka2016,Wang2014,Li2015,
Hesselmann2016,Toldin2015,Corboz2018}.
This consistency supports our argument based on the small system sizes for the present model.
The valid range of $l_B (<L_y)$ would be changed and correspondingly
results could be improved if we include numerical data for larger system sizes, although it is computationally
expensive.
It has also been shown that
a scaling analysis at zero magnetic field with an even shorter length scale
works well for a related model~\cite{Corboz2018}.  

By numerically fitting the discrete data for $B\gtrsim 0.02B_0$ in our system, one can obtain continuum curves which
smoothly connect them.
To this end,  as discussed above, 
we first observe $\varepsilon(B)- \varepsilon(0)\sim l_B^{-3}=B^{3/2}$ in the
Dirac semimetal phase, while
$\varepsilon(B)- \varepsilon(0)\sim l_B^{-4}=B^2$ in the CDW phase.
Then,
we introduce the following fitting functions so that their leading functional forms are consistent with these behaviors,
\begin{align}
\varepsilon_{\rm fit}(B)=\left\{
\begin{array}{ll}
a_0+a_1l_B^{-3}+a_2l_B^{-4} & (V\leq V_c), \\
a_0+a_1l_B^{-4}+a_2l_B^{-5} & (V> V_c),
\end{array}
\right.
\label{eq:Efit}
\end{align}
where $a_j$ are fitting parameters. We have also included the higher order terms.
We note that the zero-field energy $a_0$ is robust to the fitting 
even when we include further higher order terms in $l_B^{-1}$. 

Given the extrapolated
ground state energy density $\varepsilon$, we can now evaluate the orbital magnetization,
\begin{align}
M(V,B)=-\frac{\partial \varepsilon(V,B)}{\partial B}.
\label{eq:M}
\end{align}
Although the magnetic field has to be a continuum variable in this formula,
it is discrete $B=n\times \delta B$ in our calculations and we find that numerical differentiation
$\delta \varepsilon/\delta B$
is not so reliable as will be seen in the following.
Therefore, 
we mainly focus on the fitting function $\varepsilon_{\rm fit}(B)$
and differentiate it analytically to obtain the magnetic moment $M$.

\begin{figure}[tbh]
\includegraphics[width=6.5cm]{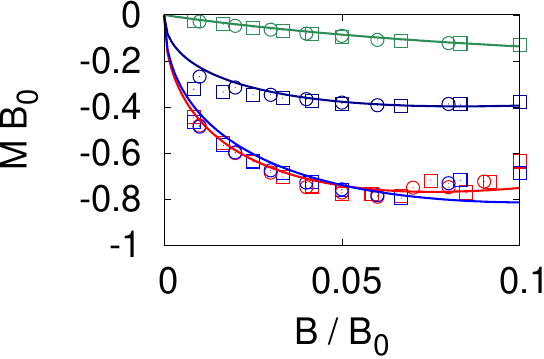}
\caption{The orbital diamagnetic moment where the parameters are same as in Fig.~\ref{fig:eB}.
The symbols are forward differentiation of the calculated data, 
while the solid curves are analytic differentiation of $\varepsilon_{\rm fit}(B)$.
}
\label{fig:MB}
\end{figure}

In Fig.~\ref{fig:MB}. 
we show the results obtained from the fitting function $\varepsilon_{\rm fit}(B)$ (solid curves),
and also the direct forward differentiation of the calculated discrete data with symbols for a comparison.
For weak interactions,
we clearly see that $M(V=0)$ and $M(V=0.5t)$ are very close each other,
and think that the small difference is not so physically relevant as will be revisited later.
The robustness of $M(V)$ for small $V$ is understood as a result of non-trivial cancellations of two opposite effects.
Firstly, at $B=0$, the Fermi velocity $v_F(V)$ 
is increased by $V$ according to the recent numerical studies
of the $t$-$V$ model~\cite{Schuler2021} and it remains regular even at the critical point 
according to the scaling analysis~\cite{Igor2009vF}, $v_F\sim (V_c-V)^{\nu(z-1)}$ with 
the dynamical critical exponent $z=1$.
In a simple Fermi liquid picture around the non-interacting limit,
the orbital magnetization of Dirac fermions is expected to be renormalized roughly as
$M(V)/M(0)\propto v_F(V)/v_F(0)$.
At the same time, however, the magnetic catalysis generating fermion mass $\propto B$ at weak interactions
~\cite{Shovkovy2013book,Miransky2015review,Fukushima2019review,
Gusynin1994,Gusynin1996NPB,Tada2020}
will suppress the magnetization $M$.
As a result of the non-trivial cancellation between these two effects, $M$ can remain almost unchanged for small $V$.
Furthermore, we aruge that
the near constant $M(V,B)$ in presence of the weak interaction $V$ is 
not  specific to the present model Eq.~\eqref{eq:H} and it is
a general property of Dirac systems
whose criticality belongs to the chiral Ising universality class, as will be discussed later based on 
a scaling analysis. 
This can be compared with the previous results for long-range Coulomb interacting Dirac electrons
where the magnetic catalysis is dominant over the Fermi velocity enhancement at zero temperature and 
consequently the diamagnetism is suppressed~\cite{Yan2017}.
In addition, suppression of diamagnetism is expected to occur in other correlated Dirac fermions 
where the Fermi velocities
are decreased by the interactions~\cite{Tang2018,Hesselmann_com2019,Tang_rep2019,Lang2019,Unozawa2020}.
In Fig.~\ref{fig:MB},
as the magnetic field becomes stronger, 
$|M(V=0.5t)|$ becomes even larger than $|M(V=0)|$ within the present model calculation.
This behavior is related to the subleading terms in $\varepsilon_{\rm fit}(B)$ and 
it seems to be a non-universal, model-dependent property at least in Fig.~\ref{fig:MB}.
This point will also be revisited later.

When the interaction becomes stronger $V\gtrsim V_c$,
the $B$-dependence of the energy density $\varepsilon$ gets weaker,
which means that the orbital magnetic moment $M$ is simply suppressed by the interaction $V$,
as seen in Fig.~\ref{fig:MB}. 
By increasing the interaction, the magnetization $M$ decreases monotonically with the qaulitative change
from $M(V\leq V_c,B)\sim \sqrt{B}$ to $M(V> V_c,B)\sim B$.
We see that $M\sim B$ indeed holds in the direct numerical differentiation of the discrete data
and they agree well with the fitting result.
As the interaction increases further, $V\to\infty$, the Dirac mass becomes larger and 
finally the orbital magnetization approaches zero, $M\to0$.

To elucidate universal aspects of the diamagnetism in the present Dirac system whose criticality belongs to the
chiral Ising universality class,
we now introduce a scaling ansatz for the singular part of the ground state energy density
in the thermodynamic limit $L_y\to\infty$, 
\begin{align}
\varepsilon_{\rm sing}(g,l_B^{-1})=b^{-D}\varepsilon_{\rm sing}(b^{y_g}g,bl_B^{-1}),
\label{eq:ansatz}
\end{align}
where $g$ is the reduced interaction $g=(V-V_c)/V_c$
~\cite{Tada2020,Fisher1991,Lawrie1997,Tesanovic1999}. 
The scaling dimension of $g$ is $y_g=1/\nu$ with the correlation length exponent $\xi\sim |g|^{-\nu}$,
and the dimensionality is $D=2+z=3$ 
with the dynamical critical exponent $z=1$ for the present Lorentz symmetric criticality.
This scaling ansatz can describe the critical behaviors around the QCP
as a function of $B$,
which belongs to the chiral Ising universality class with four component Dirac fermions in the present case.
The proposed scaling ansatz is formally similar to the conventional finite size scaling ansatz for
the isotropic system size $L$ in absence of the magnetic field, 
$\varepsilon_{\rm sing}(g,L^{-1})=b^{-D}\varepsilon_{\rm sing}(b^{y_g}g,bL^{-1})$.
These two ansatzes are related through the energy density at non-zero $l_B^{-1}, L^{-1}$: 
We have the ansatz Eq.\eqref{eq:ansatz} for $l_B\ll L\to\infty$, while the conventional one is obtained for 
$l_B\to\infty\gg L$.
In the previous study on the same model Eq.~\eqref{eq:H}~\cite{Tada2020},
we have shown that the scaling ansatz similar to Eq.~\eqref{eq:ansatz} indeed holds and 
obtained the critical exponents $\nu=0.80(2), \beta=0.54(3)$, and  the critical interaction strength
$V_c=1.30(2)t$, where $\beta$
is the CDW order parameter exponent
$M_{\rm CDW}\sim g^{\beta}$ for $g\geq0$.
These values are consistent with those obtained in other studies at zero magnetic field
~\cite{Sorella1992,Assaad2013,Otsuka2016,Wang2014,Li2015,
Hesselmann2016,Toldin2015,Corboz2018}.
In the present study, we simply use these previous results and examine quantum criticality of 
the orbital magnetization.

From the scaling ansatz Eq.\eqref{eq:ansatz} with $b=l_B$,
the total ground state energy density is regarded as a function of the single variable $gl_B^{1/\nu}$
with the trivial $l_B^{-D}$ factor,
\begin{align}
\varepsilon(g,l_B^{-1})=\varepsilon_0(g)+\frac{\Phi(gl_B^{1/\nu})}{l_B^D}(1+cl_B^{-\omega})+\cdots.
\label{eq:E}
\end{align}
We have included a correction to scaling to improve the scaling description,
and 
similar corrections with respect to the system size $L$ have been often used in numerical calculations
~\cite{Otsuka2016},
although physical origin of the introduced corrections may not be so clear in general.
In this study,
we regard the correction to scaling as a working ansatz to evaluate large $l_B$ behaviors in a systematic way. 
The scaling function $\Phi(x)$ is universal in the sense that it is determined only by the universality class, 
and it should be independent of boundary conditions of the system since Eq.~\eqref{eq:E} is the energy density
in the thermodynamic limit.
This is in sharp contrast to the conventional finite size corrections which depends on boundary conditions.
Note that the universal function $\Phi(x)$ behaves as $\Phi(x\ll -1)\sim $ const. corresponding to
$\varepsilon-\varepsilon_0\sim l_B^{-3}$ in the Dirac semimetal phase for $g<0$ , while
$\Phi(x\gg 1)\sim x^{-\nu}$ corresponding to $\varepsilon-\varepsilon_0\sim l_B^{-4}$ in the CDW phase for $g>0$.
Around the QCP, $\Phi(x)$ should be analytic in $x$ since there would be no phase transitions for 
any nonzero $l_B^{-1}$, and $\Phi(0)$ at $g=0$ may contain some useful information about the criticality
as will be discussed later.

\begin{figure}[tbh]
\includegraphics[width=6.5cm]{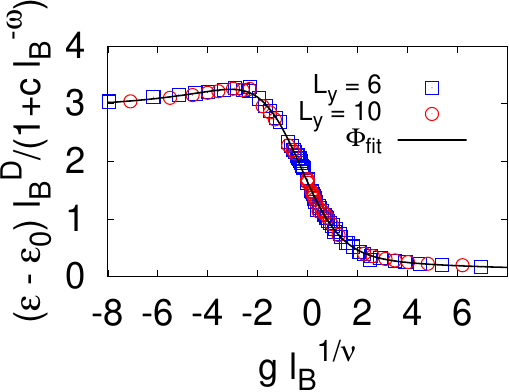}
\caption{The scaling plot of the ground state energy density $\varepsilon(g,l_B^{-1})$ for 
$L_y=6$ (squares) and $L_y=10$ (circles)
with the fitting function $\Phi_{\rm fit}(gl_B^{1/\nu})$ (black solid curve).
The interaction is $V=0.50t\sim 2.0t$ and
the magnetic length is $l_B\simeq 3.2l_{B_0}\sim7.1l_{B_0}$.
}
\label{fig:e_scaling}
\end{figure}

To show a scaling plot of $\varepsilon(g,l_B^{-1})$, 
we use $\varepsilon_0(g)=a_0(g)$ from Eq.~\eqref{eq:Efit} which are robust to details of the fitting.
Then, the calculated $\varepsilon$ collapse onto a single curve as shown in Fig.~\ref{fig:e_scaling}
with the critical exponent $\nu=0.80$ and critical interaction $V_c=1.30t$~\cite{Tada2020}.
Here, the interaction range is relatively wide, $V=0.50t\sim 2.0t$, and
the magnetic length is $l_B\simeq 3.2l_{B_0}\sim7.1l_{B_0}$ measured in unit of $l_{B_0}=1/\sqrt{2\pi}$.
The overall behavior of $\Phi_{\rm data}(gl_B^{1/\nu})\equiv(\varepsilon-\varepsilon_0)l_B^3/(1+cl_B^{-\omega})$ 
is consistent with the above mentioned general expectation.
Based on this observation,
we introduce the following fitting function as a working ansatz,
\begin{align}
\Phi_{\rm fit}(x)=\alpha_0+\alpha_1\tanh [\alpha_2(x-\alpha_3)]
+\frac{\alpha_4}{[(x-\alpha_5)^2+\alpha_6]^{\nu/2}},
\label{eq:Phi}
\end{align}
where $\alpha_0=-\alpha_1$ and $\alpha_j$ are parameters to be determined from numerical fitting with the calculated data (see Appendix~\ref{appB} for details).
The solid curve in Fig.~\ref{fig:e_scaling} is the fitting function $\Phi_{\rm fit}(x)$ and it well agrees with the data 
(variance of residuals $\chi^2= O(10^{-3}$)).

\begin{figure}[tbh]
\includegraphics[width=6.5cm]{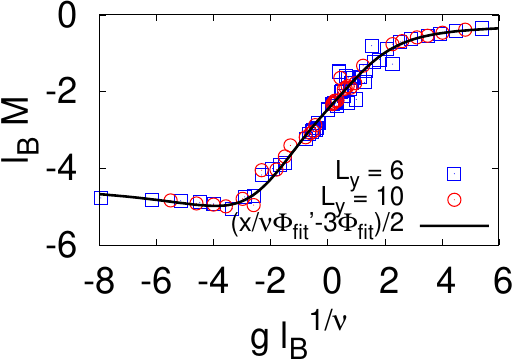}
\caption{The scaling plot of the orbital magnetization $l_BM$ as a function of $x=g\l_B^{1/\nu}$.
The solid curve is calculated from the fitting function $\Phi_{\rm fit}(x)$.
The symbols are calculated directly from the numerical data for $L_y=6$ (squares) and $L_y=10$ (circles)
with forward differentiation.
}
\label{fig:m_scaling}
\end{figure}

Once the scaling function has been obtained,
we can find the universal scaling of the orbital magnetization $M=-\partial \varepsilon/\partial B
=(1/2l_B^{3})\partial (\Phi l_B^{-3})/\partial l_B$ with suppressing the non-universal correction term
near the QCP for sufficiently large $l_B$,
\begin{align}
l_B{M}
=\frac{1}{2}\left( \frac{gl_B^{1/\nu}}{\nu}\Phi'(gl_B^{1/\nu})-D\Phi(gl_B^{1/\nu})\right).
\label{eq:Muni}
\end{align}
This equation clearly means that the orbital magnetization in the form 
${\mathcal M}(gl_B^{1/\nu})\equiv l_BM(V,B)$ is a universal function of $gl_B^{1/\nu}$ characteristic of
the associated quantum criticality, namely, the $N=4$ chiral Ising universality class in $D=(2+1)$-dimensions.
We show ${\mathcal M}$ obtained from $\Phi_{\rm fit}$ in Fig.~\ref{fig:m_scaling}.
For a comparison, 
we also show the results calculated with forward differentiation of the numerical data.
Although the numerical differentiation of our data is less accurate due to its discreteness,
overall behaviors are in agreement with the one obtained from the analytic differentiation of $\Phi_{\rm fit}(x)$.
As explained above, the scaling function behaves as $\Phi(x\ll-1)\sim {\rm const}$ and therefore
we have $M\sim -l_B^{-1}\sim -\sqrt{B}$ in the Dirac semimtal phase.
Similarly, $\Phi(x\gg1)\sim x^{-\nu}$ implies
$l_BM\sim -(gl_B^{1/\nu})^{-\nu}\propto -l_B^{-1}$,
which means $M\sim -B$ in the CDW phase as expected.
At the QCP, the magnetization is $M=-(3/2)\Phi(0)\sqrt{B}$, where the amplitude is expected to be
universal as will be discussed in the next section.

\begin{figure}[tbh]
\includegraphics[width=8.5cm]{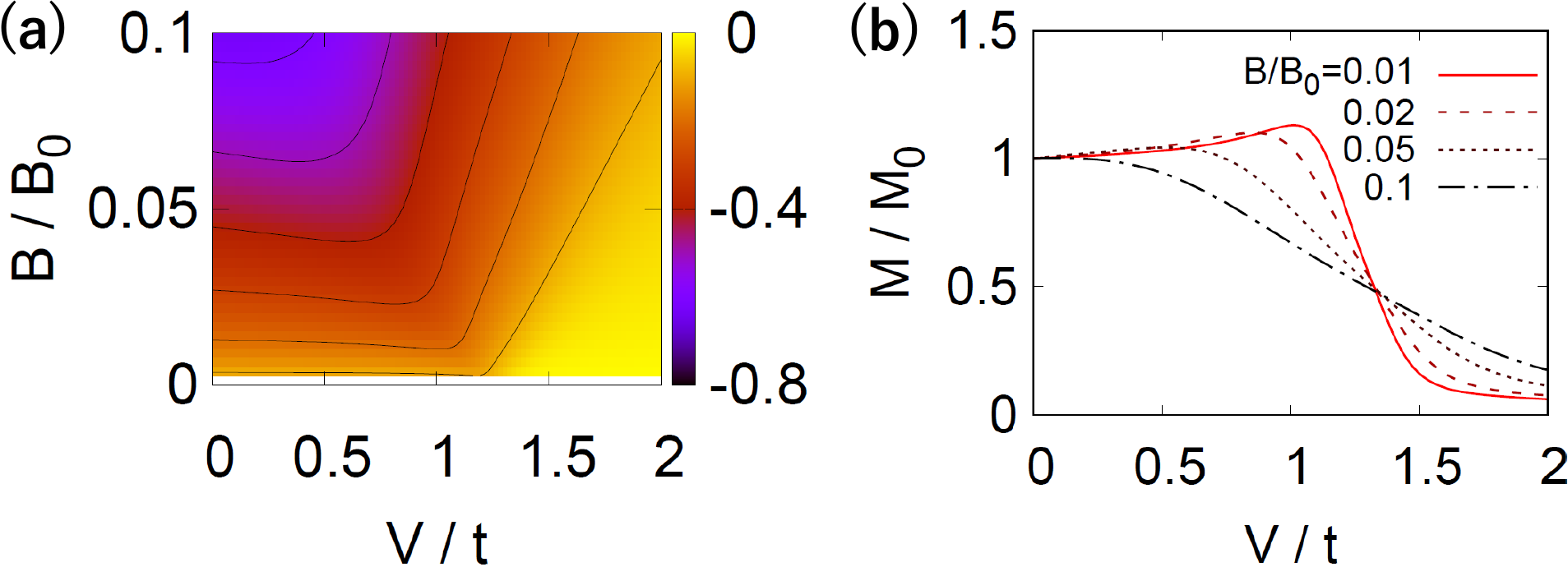}
\caption{(a) The orbital magnetization calculated as $M(V,B)=l_B^{-1}{\mathcal M}(gl_B^{1/\nu})$
in the $V$-$B$ plane and (b) $M$ as a function of $V$ normalized by 
$M_0=M(V=0,B)$ for several values of $B$.
The QCP is located at $V_c=1.30t$.
The contour curves in (a) are at $M=-0.01\sim-0.06$ with the uniform interval 0.01.
}
\label{fig:M2d}
\end{figure}

Now we revisit $M(V,B)$ as a function of $V$ and $B$ with using the scaling function,
$M(V,B)=l_B^{-1}{\mathcal M}(gl_B^{1/\nu})$.
Although $M(V,B)$ has already been shown in Fig.~\ref{fig:MB}, there were non-universal finite $l_B$ corrections
and such corrections can be removed with use of ${\mathcal M}(x)$.
Here, we simply assume that ${\mathcal M}(x)$ is applicable for all $-\infty<x<\infty$,
although it is more reliable for a small $x=(V/V_c-1)l_B^{1/\nu}$ region.
Thus the following discussions can elucidate universal aspects of the orbital magnetization $M$
which are so clear in Fig.~\ref{fig:MB}.
We show $M(V,B)=l_B^{-1}{\mathcal M}(gl_B^{1/\nu})$ in Fig.~\ref{fig:M2d}.
Since the scaling function ${\mathcal M}$ has three distinct regimes, the magnetization $M$ in the $V$-$B$ plane
shows corresponding behaviors respectively for $gl_B^{1/\nu}\ll-1$ (``Dirac semimetal regime"),
$|gl_B^{1/\nu}|\ll1$ (``quantum critical regime"), and $gl_B^{1/\nu}\gg1$ (``CDW regime").

In the Dirac semimetal regime corresponding to the left-bottom region in the $V$-$B$ plane of Fig.~\ref{fig:M2d} (a), 
the diamagnetism is highly robust to the interaction.
For example at a small magnetic field, $B/B_0=0.01$, 
the orbital magnetization is robust up to the interaction $V/t\simeq 1.0$ and then sharply
drops in the quantum critical regime around the QCP, $V_c=1.30t$, as seen in Fig.~\ref{fig:M2d} (b).
Such a behavior is commonly seen for other small values of $B$ 
and $M(V)$ as a function of $V$ gets smeared for larger values of $B$.
In the CDW regime, $M$ is strongly suppressed by the interaction.
The crossover lines separating the different regimes 
are roughly given by $|gl_B^{1/\nu}|\sim 1$ or equivalently $\xi\simeq l_B$, 
namely, $|V-V_c|^{\nu}\sim \sqrt{B}$.
In addition, by looking at the magnetization $M$ closely,
one can see that $M$ is slightly enhanced by the interaction $V$ at small magnetic fields,
and it is free from non-universal (model dependent) finite size corrections in contrast to Fig.~\ref{fig:MB}. 
We stress that the present argument is based on the scaling ansatz, and therefore
the resulting robust and slightly enhanced orbital magnetization in presence of the interaction $V$
should be a common property in general Dirac systems 
whose criticality belongs to the chiral Ising universal class as long as the ansatz holds.
As already mentioned, the robust $M$ in presence of the interaction $V$ is a consequnce
of the non-trivial competition between the renomalized Fermi velocity and magnetic catalysis.
The present results imply that the former effect is dominant over the latter when $V$ is weak in the
chiral Ising universality class.
This is contrasted with the diamagnetism in 
the Hubbard models and also in systems with
the long-range Coulomb interaction, 
where the Fermi velocity renormalization cannot cancel out effects of the magnetic catalysis~\cite{Tang2018,Hesselmann_com2019,Tang_rep2019,Lang2019,Yan2017}.

\section{Discussion and Summary}
\label{sec:summary}

\begin{figure}[tbh]
\includegraphics[width=6.5cm]{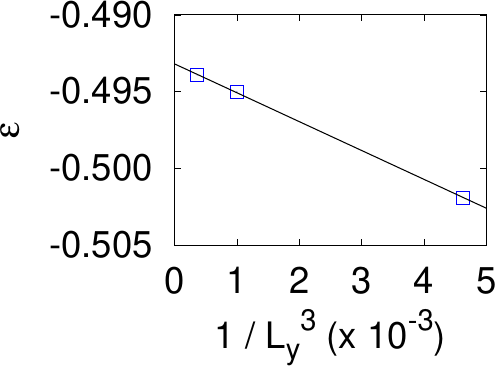}
\caption{The ground state energy density at $V=V_c$ without a magnetic field, 
$\varepsilon(g=0,l_B^{-1}=0,L_y^{-1})$, for system sizes $L_y=6, 10, 14$ with the periodic boundary
condition for $y$-direction.
The qualitative behavior is assumed to be $\varepsilon-\varepsilon_0\sim 1/L_y^3$, as indicated by the solid line. 
}
\label{fig:eL}
\end{figure}

As menioned before, the scaling behavior Eq.~\eqref{eq:E} is seemingly similar to the conventional
finite system size scaling at zero magnetic field, 
$\varepsilon(g,L^{-1})=\varepsilon_0(g)+\tilde{\Phi}(gL^{1/\nu})/L^D+\cdots$.
The leading finite size correction $\tilde{\Phi}(0)/L^D$ is called the Casimir energy density
in field theories and contains universal information of the criticality.
In $D=1+1$ dimensions, the Casimir amplitude is written as $\tilde{\Phi}(0)\sim cv$ with
a boundary condition dependent coefficient, where $v$ is the speed of light (velocity of excitations)
and $c$ is the central charge of the underlying conformal field theory~\cite{Cardy,CFT_Casimir1,CFT_Casimir2}.
Generalizations to higher dimensional systems have been first discussed for 
a cylinderical space-time geometry~\cite{Cardy}
and also recently examined in torus and infinite systems~\cite{Schuler2016,Rader2018,Schuler2021}.
It was proposed that the Casimir amplitude in a $(2+1)$-dimensional torus system is
decomposed as $\tilde{\Phi}_{\rm torus}(0)=C_{\rm torus}v$, where $C_{\rm torus}$ contains some universal information
of the underlying field theory.
For a comparison,
we also calculate the Casimir energy density in our model as shown in Fig.~\ref{fig:eL},
where the ground state energy density is assumed to behave as
$\varepsilon(g=0,l_B^{-1}=0,L_y^{-1})=
\varepsilon_0(0)+\tilde{\Phi}_{\rm iDMRG}(0)/L_y^D+\cdots$,
as in other Lorentz symmetric critical systems~\cite{Schuler2016,Rader2018,Schuler2021}. 
The amplitude is found to be negative $\tilde{\Phi}_{\rm iDMRG}(0)<0$ in contrast to 
$\Phi(0)>0$ in Eq.~\eqref{eq:E}.
Within infinite projected entangled pair states (iPEPS) calculations,
the correlation length 
$\xi_D$ due to a finite bond dimension 
can be a new length cut-off scale in a thermodynamically large system and
will play a similar role to that of the system size $L$,
leading to $\varepsilon=\varepsilon_0+C_{\rm iPEPS}v/\xi_D^{^3}+\cdots$ at $g=0$~\cite{Rader2018}.

We have demonstrated in this study that,
in presence of a magnetic field,
the leading term $\Phi(0)/l_B^3$ in a thermodynamically large system can be regarded as a magnetic analogue
of the conventional Casimir energy in a finite size system, and may be called {\it magnetic Casimir energy}.
Note that similarity between conventional Casimir energy and magnetic Casimir energy is already 
implied in single-particle spectra of the non-interacting Dirac electrons;
$\epsilon_{}(L)\propto v_F/L$ in a finite size system without 
a magnetic field,
and 
$\epsilon(l_B)\propto v_F/l_{B}$ in an infinite system with a magnetic field.
Structures of the single-particle spectrum are governed by the Lorentz symmetry and 
the characteristic length scale is either $L$ or $l_B$.
These properties are also common to a correlated Dirac system around a QCP,
leading to similar functional forms of the critical Casimir energies.
However, there is an essential difference between them; 
the conventional Casimir energy is geometry (boundary condition) dependent,
while the magnetic Casimir energy is independent of boundary conditions since it is the energy in the thermodynamic
limit. 
Besides, the magnetic Casimir energy can be controlled by an external magnetic field,
which is a difference from the Casimir energy in iPEPS for an infinite system
that is a purely theoretical quantity.

It is known that the Casimir energy $L^2\tilde{\Phi}(gL^{1/\nu})/L^D$ leads to 
the critical Casimir force $f_L=-\partial E/\partial L, E=L^2\varepsilon$, 
and related physics has been extensively studied in various systems
which exhibit finite temperature {\it classical} phase transitions
~\cite{FdG1978,Garcia1999,Fukuto2005,Hertlein2008,
Hucht2007,Vasilyev2009,Gambassi2009,Hucht2011,Hasenbush2012,Krech,Gambassi2009review}.
The universal nature of the classical critical Casimir force has been experimentally observed 
for example in a binary liquid mixture in thin film geometry~\cite{,Garcia1999,Fukuto2005,Hertlein2008}.
The quantum critical orbital diamagnetism discussed in this study can be regarded as a magnetic, quantum 
analouge of the classical critical Casimir force.
Indeed, the magnetization is rewritten as
$L^2M=-\partial E/\partial B=-f_B/2l_B^3$ with $f_B=-\partial E/\partial l_B$,
and the ``effective force" $f_B$ is repulsive for diamagnetism. 
This is contrasting to the attractive real space Casimir force in our model (Fig.~\ref{fig:eL}),
corresponding to the sign difference between $\Phi(0)$ and $\tilde{\Phi}_{\rm iDMRG}(0)$.
The repulsiveness of the effective force $f_B$ could be compared with general Casimir forces,
where they are usually attractive and repulsive forces are rarely realized~\cite{Munday2009,Jiang2019}.

\begin{figure}[tbh]
\includegraphics[width=8.5cm]{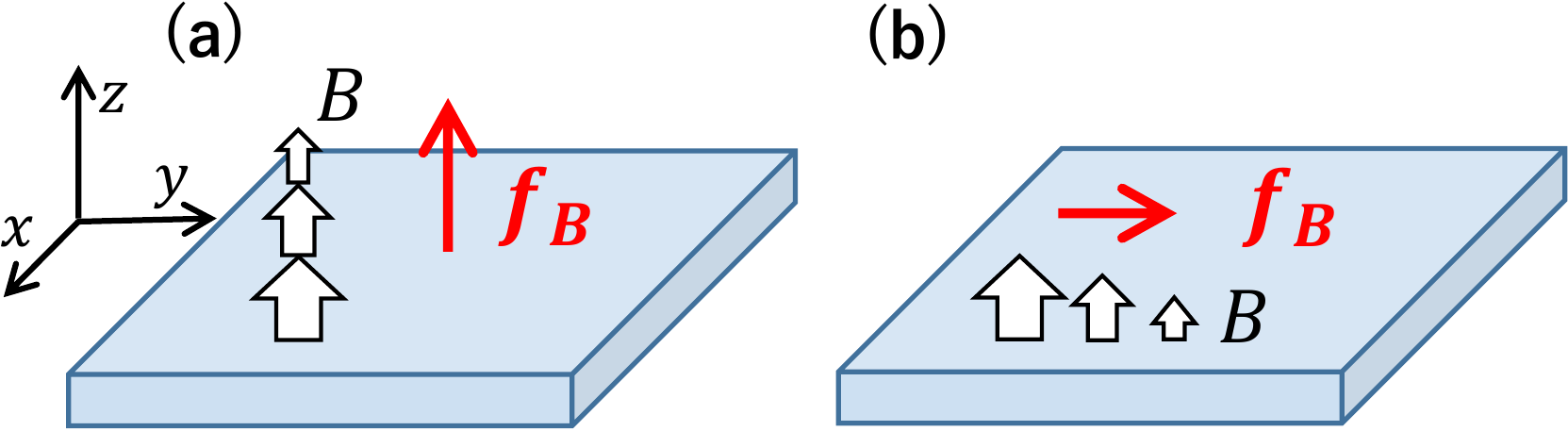}
\caption{(a) Schematic picture of the effective Casimir force $f_B$ when the $z$-axis magnetic field $B$
macroscopically depends on the position $z$ and (b) varies in the $xy$-plane.
If $B(z)$ is decreasing ($l_B(z)$ is increasing) with $z$,
the effective force is repulsive $f_B>0$ (or equivalently, the real force is $f_z=-\partial E/\partial z>0$). 
Similary, for $B(y)$ decreasing in the $y$-direction, the effective force will be $f_B\parallel +\hat{y}$
(or $f_y=-\partial E/\partial y>0$).
}
\label{fig:force}
\end{figure}

Interestingly, the critical orbital magnetization $M$ or equivalently the effective force $f_B$ 
could be measured in experiments by carefully controlling experimental
parameters, 
which may be advantageous over the formidable challenge for 
a direct observation of the real space Casimir force in a solid crystal.
In a bulk magnetization measurement, observing the magnetization $M$ is just equivalent to measuring $f_B$,
and one can understand the above analogy in a visible manner.
For example,
the effective force $f_B$ could be measured by the conventional magnetization measurement with
the Faraday balance, where a mechanical force $f_z(z)$ under a macroscopically
non-uniform magnetic field $B(z)$ is observed
in the typical setup shown in Fig.~\ref{fig:force} (a).
The effective force is related with the mechanical force simply as
$f_z=-\partial E/\partial z=f_B\cdot \partial l_B/\partial z$.
Besides, note that the direction of $f_B$ is determined by the macroscopic configuration of the magnetic field,
since the energy density $\varepsilon(l_B)$ of a diamagnetic Dirac system favors a larger $l_B$ (a smaller $B$).
If the $z$-axis magnetic field varies in the $xy$-plane in a macroscopic length scale
as shown in Fig.~\ref{fig:force} (b),
the effective force $f_B$ will be parallel to the plane. 
Such an in-plane force would be more analogous to the critical Casimir force $f_L$ which also acts within the plane, 
but the system may exhibit
very complex behaviors in an inhomogeneous setup.

In summary, we have discussed orbital diamagnetism in correlated Dirac electrons with 
use of iDMRG for the $t$-$V$ lattice model which exhibits the CDW quantum phase transition.
The orbital diamagnetism is robust to the short-range interaction $V$ in the Dirac semimetal regime, 
while it is suppressed for a strong $V$ in the CDW regime.
The robustness of the diamagnetism to the interaction
is understood as  a consequence of non-trivial competition between the enhanced Fermi velocity
and mass generation by the magnetic catalysis.
Furthermore, it is concluded that the robust orbital diamagnetism is a universal property of Dirac systems in the
chiral Ising universality class based on the scaling analysis in terms of magnetic length.
The analogy between the quantum critical diamagnetism and critical Casimir effect was
discussed.
To our best knowledge, this is a first unbiased numerical calculation of the orbital diamagnetism in
strongly interacting electrons other than one-dimensional systems
~\cite{Orignac2001,Carr2006,Roux2007,Greschner2015,Buser2021}, 
and it could provide a basis for further theoretical developments
in this field.

\section*{acknowledgements}

We thank F. Pollmann for introducing TeNPy to us.
The numerical calculations were done at the Max Planck Institute for the Physics of Complex Systems.
This work was supported by JSPS KAKENHI Grant No. JP17K14333 and
by a Grant-in-Aid for Program for Advancing
Strategic International Networks to Accelerate the Circulation
of Talented Researchers (Grant No. R2604) ``TopoNet."

\appendix
\section{Extrapolation of bond dimension}
\label{app}
\begin{figure}[tbh]
\includegraphics[width=6.5cm]{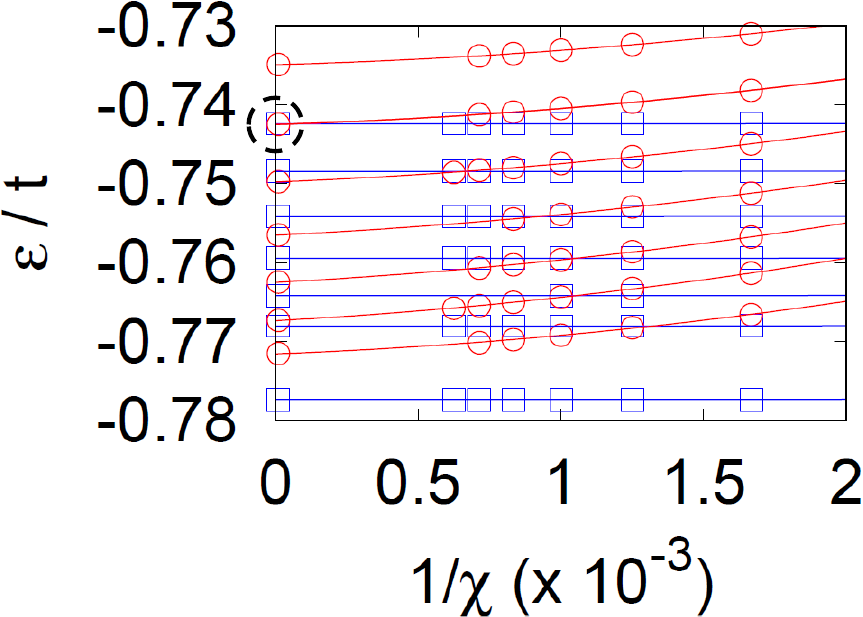}
\caption{Extrapolation of the ground state energy density $\varepsilon$ for the $\chi\to\infty$ limit
at $V=0.5t$.
The blue (red) symbols are for $L_y=6 (L_y=10)$, and the curves are polynomial fittings.
The corresponding magnetic fields for each curve are $B/\delta B=0,1,2,3,4,5,6$ from the bottom to the top,
where
$\delta B=2\pi/(6\times 20)$ for $L_y=6, L_x'=20$ and $\delta B=2\pi/(10\times 10)$ for $L_y=10,L_x'=10$. 
As marked by a broken circle,
at $B=2\pi/20=2\pi\times 6/(6\times20)=2\pi\times 5/(10\times10)$,
the extrapolated $\varepsilon(\chi\to\infty)$ for two system sizes coincide within the error (which
is smaller than the symbols).
On the other hand, $\varepsilon(\chi\to\infty)$ for $L_y=6,10$ differ at $B=0$ because 
the corresponding magnetic length is $l_B=\infty\gg L_y$.
}
\label{fig:e_ext}
\end{figure}

All the calculation results in the main text are obtained by the extrapolation to $\chi\to\infty$
from the finite bond dimensions up to $\chi=1600$.
For example, we show in Fig.~\ref{fig:e_ext} 
the ground state energy density at $V=0.5t$ for two different system sizes $L_y=6,10$
obtained by finite $\chi$ calculations together with polynomial fitting curves.
We find that the extrapolation works well in the present model as mentioned before,
and standard deviations of the extrapolated $\varepsilon$ are less than 0.01\% and 
are smaller than the symbols in Fig.~\ref{fig:e_ext}, which is sufficient for
the purpose of the present study.
It is confirmed that other extrapolation schemes such as the
linear fitting with respect to the truncation error give consistent results.
The $\chi$-extrapolation was used also for the CDW order parameter in the previous study~\cite{Tada2020}
and is employed in the present study as well, which enables us to discuss two studies in a coherent manner.
We have performed extrapolations as in Fig.~\ref{fig:e_ext} for all other parameter values, 
and consider only the extrapolated energy density $\varepsilon(\chi\to\infty)$ in our discussion.

\section{Details of fitting with Eq.~\eqref{eq:Phi}}
\label{appB}
\begin{figure}[tbh]
\includegraphics[width=6.5cm]{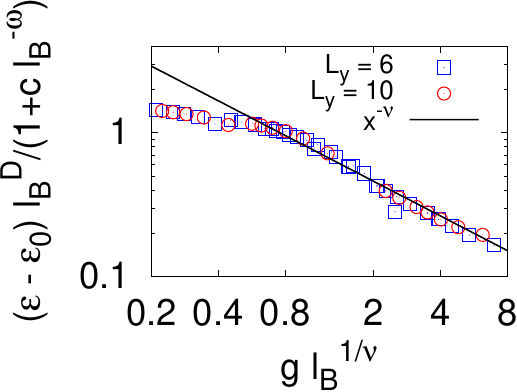}
\caption{The scaling plot of the ground state energy density $\varepsilon(g,l_B^{-1})$ 
at large $gl_B^{1/\nu}$ for 
$L_y=6$ (squares) and $L_y=10$ (circles).
The solid line represents the qualitative behavior $\sim x^{-\nu}$ with $\nu=0.80$.
}
\label{fig:e_scaling_lim}
\end{figure}

The numerically obtained scaling function $\Phi_{\rm data}(\cdot)$ is consistent with the limiting behaviors of
the ground state energy density at $x\to \pm\infty$.
For example, $\Phi_{\rm data}(x\gg1)\sim x^{-\nu}$ can indeed be confirmed in Fig.~\ref{fig:e_scaling_lim}.
We also find that $\Phi_{\rm data}(x\ll -1)$ is roughly 
$\Phi_{\rm data}(x)-\Phi_{\rm data}(-\infty)\sim |x|^{-\nu}$ (not shown),
which leads to a natural behavior, $\varepsilon= \varepsilon_0
+{\rm const}\times l_B^{-3}+{\rm const}\times l_B^{-4}+\cdots $ in the Dirac semimetal phase.
These observations enable us to evaluate the universal scaling function $\Phi(\cdot)$ in Eq.~\eqref{eq:E} 
by using simple known functions and to introduce the fitting function $\Phi_{\rm fit}(\cdot)$ in Eq.~\eqref{eq:Phi}.
We note that it is not trivial to have
a successful scaling plot over a wide range of $x=gl_B^{1/\nu}$
where $\Phi(x)$ does not have a simple Taylor expansion with a small order in $x$.
However, such a non-trivial scaling has been often examined in classical statistical models
for the Casimir effect~\cite{FdG1978,Garcia1999,Fukuto2005,Hertlein2008,
Hucht2007,Vasilyev2009,Gambassi2009,Hucht2011,Hasenbush2012,Krech,Gambassi2009review}.
The successful description of scaling behaviors of the ground state energy density
consequently suggests that the scaling ansatz Eq.~\eqref{eq:E} is indeed correct,
which is a priori non-trivial.
Together with the previous scaling argument for the CDW order parameter~\cite{Tada2020}, 
the present study supports the magnetic length scaling ansatz for which there are only few studies
~\cite{Fisher1991,Lawrie1997,Tesanovic1999}.

\bibliography{ref}

\end{document}